\documentclass{article}

\usepackage{arxiv}

\usepackage[utf8]{inputenc} 
\usepackage[T1]{fontenc}    
\usepackage{hyperref}       
\usepackage{url}            
\usepackage{booktabs}       
\usepackage{amsfonts}       
\usepackage{nicefrac}       
\usepackage{microtype}      
\usepackage{lipsum}
\usepackage{graphicx}
\usepackage{enumitem}
\usepackage{gensymb}
\usepackage{float}

\usepackage{amsmath}
\usepackage{subcaption}
\usepackage{physics}
\usepackage{tabularray}

\usepackage{xcolor}
\definecolor{custompink}{HTML}{FB6F92}
\definecolor{customgray}{HTML}{dad7cd}

\usepackage{amsthm, amssymb, thmtools}

\usepackage{mathtools}

\makeatletter
\newcommand*{\textlabel}[2]{%
  \edef\@currentlabel{#1}
  \phantomsection
  #1\label{#2}
}
\makeatother

\usepackage[ruled,linesnumbered]{algorithm2e}
\makeatletter
\newcommand{\nosemic}{\renewcommand{\@endalgocfline}{\relax}}
\newcommand{\dosemic}{\renewcommand{\@endalgocfline}{\algocf@endline}}
\let\oldnl\nl
\newcommand{\nonl}{\renewcommand{\nl}{\let\nl\oldnl}}
\makeatother

\newcommand{\placeholdercite}[1]{\textbf{[REF]}}
\newcommand{\placeholderref}[1]{\textbf{[XXX]}}
\newcommand{\placeholder}[1]{\textbf{[PLACEHOLDER]}}

\usepackage[style=numeric-comp, sorting=none]{biblatex}
\title{Image-based physical characterization of \\
magnetotactic bacteria from an environmental sample}

\author{
  Mara Smite\\
  MMML lab, Department of Physics\\
  University of Latvia\\
  Riga, Latvia, Jelgavas 3, 1004 \\
  \texttt{mara.smite@lu.lv} \\
  \And
  Mihails Birjukovs\\
  MMML lab, Department of Physics\\
  University of Latvia\\
  Riga, Latvia, Jelgavas 3, 1004 \\
  \texttt{mihails.birjukovs@lu.lv} \\
  \And
  Mila Sirinelli-Kojadinovic\\
  CEA, CNRS, BIAM\\
  Aix-Marseille Universit\'{e}\\
  Saint-Paul-lez-Durance, France, 13115
  \And
  Sandrine Grosse\\
 CEA, CNRS, BIAM\\
  Aix-Marseille Universit\'{e}\\
  Saint-Paul-lez-Durance, France, 13115
  \And
  Agnese Kokina\\
  Institute of Microbiology and Biotechnology\\
  University of Latvia\\
  Riga, Latvia, 1004
   \And
  Janis Liepins\\
  Institute of Microbiology and Biotechnology\\
  University of Latvia\\
  Riga, Latvia, 1004
    \And
  Dita Gudra\\
  Latvian Biomedical Research and Study Centre\\
  Riga, Latvia, 1067
    \And
 Megija Lunge\\
  Latvian Biomedical Research and Study Centre\\
  Riga, Latvia, 1067
   \And
 Davids Fridmanis\\
  Latvian Biomedical Research and Study Centre\\
  Riga, Latvia, 1067
    \And
  Mihaly Posfai\\
  HUN-REN-PE Environmental Mineralogy Research Group\\
  Research Institute of Biomolecular and Chemical Engineering\\
  University of Pannonia\\
   Veszpr\'{e}m, Hungary, 8200\\
    \And
  Andrejs Cebers\\
 MMML lab, Department of Physics\\
  University of Latvia \\
  Riga, Latvia, Jelgavas 3, 1004 \\
    \And
  Damien Faivre\\
 Department of Physics\\
 University of Latvia \\
  Riga, Latvia, Jelgavas 3, 1004 \\
   CEA, CNRS, BIAM\\
  Aix-Marseille Universit\'{e}\\
  Saint-Paul-lez-Durance, France, 13115\\
  \And
 Guntars Kitenbergs\\
 MMML lab, Department of Physics\\
  University of Latvia \\
  Riga, Latvia, Jelgavas 3, 1004 \\
  }

\begin{document}
\maketitle

\clearpage

\begin{abstract}

Magnetotactic bacteria (MTB) are a diverse group of microorganisms that are able to biomineralize magnetic nanoparticles. Most MTB remain uncultured, making population-level characterization from natural environments difficult. We report the discovery of a new and diverse MTB-rich site in the Ogre River, Latvia, and present an integrated approach combining 16S rRNA sequencing, transmission electron microscopy, and novel open-source, automated image-based physical methods to characterize bacteria populations within environmental samples. We introduce a pipeline for cell velocimetry using a static magnetic field and a method to classify cell populations based on their magnetic moment using a modified U-turn method where cell behavior is studied in an alternating magnetic field. This study demonstrates that our physical analysis methods provide a powerful, fast, and robust toolset for MTB population analysis in complex environmental samples.

\end{abstract}


\section{Introduction}
\label{sec:intro}

Magnetotactic bacteria (MTB) are a diverse group of microorganisms both phylogenetically and morphologically \cite{Blakemore_1975, lefevre_ecology_2013, klumpp_swimming_2019, multicellular_mmb}. They stand out from other prokaryotes due to their ability to biomineralize magnetic nanoparticles within their cells. In fact, bacteria contain magnetosomes: single-domain magnetite and/or greigite crystals, surrounded by a protein-containing phospolipidic membrane \cite{bazylinski_magnetosome_2004, muller_compass_2020}. The magnetosomes in the cells exhibit various organizations in chains or randomly \cite{hanzlik_spatial_1996, THC_1_no_magnetosome_chain, le_nagard_misalignment_2019, mtb-msr1-magnetosome-moment-misalignment-in-chains}. Magnetosomes are hypothesized to enable the passive alignment of bacteria with the Earth's magnetic field (MF) \cite{goswami_magnetotactic_2022} and to help bacteria navigate complex environments \cite{mtb_in_sediment_simul_exp}. This characteristic can also be used to control the orientation and motion of cells in arbitrary directions using an external MF \cite{erglis_dynamics_2007, birjukovs-mtb-swarm-mf-control, ClusterEmergence, mtb-tunable-self-assembly-swarms,mtb-tunable-hydrodynamics}. 

Due to the intrinsic magnetic properties that allow external control of the cellular orientation, the flagellar apparatus that drives the cellular motility, and the chemical toolbox that allows cell functionalization with drugs, MTB are being studied for their applications in various fields. These include medicine \cite{vargas_applications_2018, ren_biomedical_2023, fdez-gubieda_magnetotactic_2020,villanueva_heating_2024}, microrobotics \cite{gwisai_magnetic_2022, mirkhani_spatially_2024, mtb_as_microrobots}, environmental engineering \cite{wang_mtb_environm_applic}, and others \cite{goswami_magnetotactic_2022, reactors_review}. All known MTB are Gram negative \cite{SHIVELY2009404}, but can show considerable diversity in other aspects of their biology \cite{faivre_magnetotactic_2008}. MTB have been found in the \textit{Alpha}-, \textit{Gamma}-, \textit{Eta}- and \textit{Deltaproteobacteria} classes of the phylum \textit{Proteobacteria}, the \textit{Nitrospirae} phylum \cite{lefevre_ecology_2013,THC_1_no_magnetosome_chain} and candidate \textit{Omnitrophica} phylum \cite{kolinko_op3_phylum}. This leads to considerable variation in the properties of MTB that can be determined using microscopic methods: cell and magnetosome morphology and their magnetic response \cite{faivre_magnetotactic_2008, lefevre_divers_magn_behaviour}. MTB of various shapes, including rod, spirillum, and cocci, have been imaged \cite{lefevre_divers_magn_behaviour}. Furthermore, iron-bearing magnetosome crystals can have various shapes and can be arranged in single or multiple chains, clusters, or without order within the cell \cite{faivre_magnetotactic_2008}.

Despite their natural abundance, the vast majority of MTB remain uncultured under laboratory conditions \cite{temp_effects_cocci,le_nagard_growing_2018} and research on the diversity of magnetotactic microorganisms is ongoing \cite{mtb_diversity_brazil}. The study of MTB from environmental samples provides information on their ecological roles, impact on biogeochemical cycles, diversity, and adaptations to specific environmental conditions, as well as the discovery of new species with the potential to be cultivated in the laboratory. During the first expedition in Latvia, we explored three freshwater sites and found one location on the Ogre River where MTB were abundant. In our fieldwork in Latvia, we identified an MTB-rich freshwater location on the Ogre River. The collected samples contain an unknown mixture of MTB species as well as non-magnetic bacteria. 

To study such complex environmental samples, we have developed a set of open-source image-based physical characterization tools that enable automated analysis of environmental MTB populations. These tools form the core methodology of this study and are applicable to both environmental and cultured samples. The central components of our approach include: automated detection and trajectory reconstruction; an improved Bean model, integrated with an automated U-turn analysis framework, capable of resolving multiple coexisting MTB populations based on magnetic behavior; and a novel velocimetry method requiring only a conventional optical microscope and a permanent magnet, making it highly accessible for laboratories without specialized equipment. We present a methodology that allows environmental samples containing a mixture of species based on the magnetic properties and velocity data of the cells to be analyzed, without the need of TEM or genetic analysis, thereby expanding the toolkit for magnetotactic microbiology and ecological biophysics. This is potentially especially useful in field conditions or when a rapid characterization of a complex sample is necessary.

In this paper, we combine these automated physical methods with 16S rRNA metagenomic sequencing to explore taxonomic diversity and with transmission electron microscopy (TEM) to characterize the morphology of both cells and magnetosomes of a MTB-rich natural sample. Although one goal is to evaluate whether the results converge, the methods are complementary due to their different focus (genetic, morphological, and behavioral) and scales (population vs. single-cell). This combined approach allows us to distinguish and analyze different MTB populations within a heterogeneous environmental sample based on the physical behavior of bacteria, the diversity of species, and their ecological organization.

\section{Results}

\begin{figure}[h]
\centering
\includegraphics[width=0.9\textwidth]{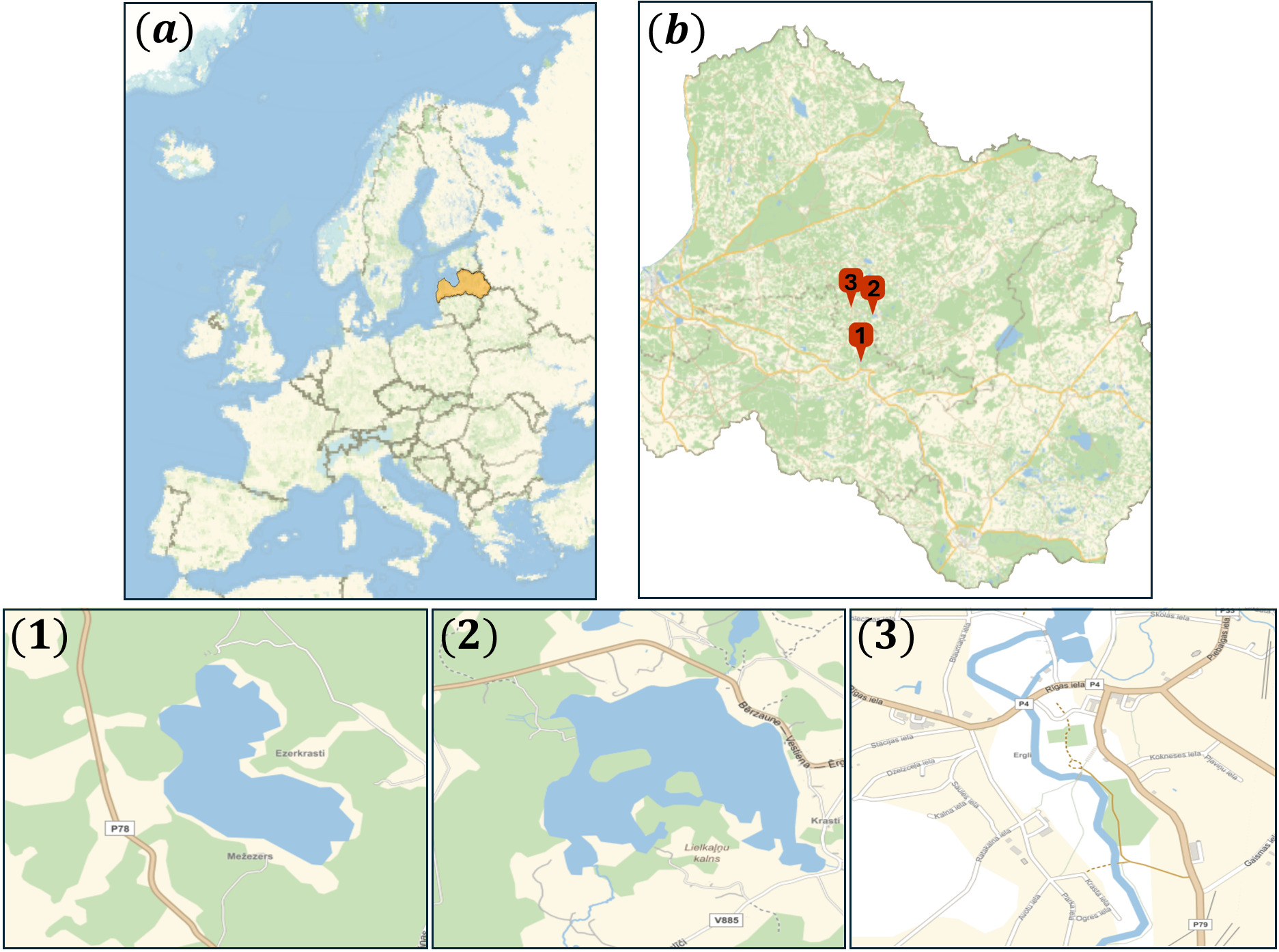} 
\caption{MTB sampling locations: a) and b) map of Latvia. Expedition A sampling locations 1 ({Lake Mezezers}), 2 ({Lake Kala}) and 3 ({Ogre River}) in search of MTB. Expedition B collected MTB from location 3 for further analysis.}
\label{fig:map}
\end{figure}

Two field expeditions were conducted in Latvia, Madona municipality (Fig. \ref{fig:map}) with the aim of identifying and characterizing MTB. \textit{Expedition A} was conducted during August 2023 with the objective of finding a source of MTB and characterizing their morphology using TEM. Three locations were sampled: Lake Mezezers (coordinates $(56.662073, 25.718520)$), Lake Kala (coordinates $(56.863128, 25.810166)$) and Ogre River (coordinates $(56.895449, 25.641971)$), with samples from the Ogre River containing abundant and morphologically different MTB. The second \textit{Expedition B} was conducted in October 2023, revisiting the site on the Ogre River, with the aim of obtaining more biomass to further characterize the diversity of MTB found at the site.

In Expedition A, a total of 24 samples were collected from three different freshwater environments. After MTB extraction, two of the containers that originated from Ogre River revealed the presence of magnetic cells, confirmed by optical microscopy. TEM images were later produced from the samples acquired in this expedition. Without enrichment, the lifespan of MTB in the sample containers did not exceed 2 days, which means that no MTB sample could be extracted from the jar after a 48-hour period. Expedition B further explored the diversity of MTB encountered in the Ogre River, yielding an MTB sample volume sufficient to perform a 16-s rRNA analysis, as well as to characterize the population using our proposed automated methods of physical cell characterization. An overview of the data collected is found in Table \ref{tab:table_dataset_overview}.


\begin{table}[H]
    \begin{center}
    \begin{tblr}{
      cells={valign=m,halign=c},
      row{1}={bg=custompink,font=\bfseries},
      colspec={|c|c|}
    }
         \hline
        \textbf{Expedition}& \textbf{Results}\\ \hline
        \textbf{A}& TEM images of cells (section: Morphological and genetic diversity section) \\ \hline
        \textbf{B}& 16S rRNA analysis (section: Morphological and genetic diversity section)\\ \hline
        \textbf{B}& Velocimetry dataset taken on \textbf{Day 1} after sampling (section: Velocimetry-based characterization)\\ \hline
        \textbf{B}& Velocimetry dataset taken on \textbf{Day 2} after sampling (section: Velocimetry-based characterization)\\ \hline
        \textbf{B}& Magnetic moment dataset (section: Magnetic moment-based characterization)\\ 
        \hline
    \end{tblr}
    \end{center}
    \caption{An overview of datasets collected during the search for MTB in Latvia.}
    \label{tab:table_dataset_overview}
\end{table}

\bigskip

\subsection*{Morphological and genetic diversity}
\label{sec:morphology}

\begin{figure}[h]
\centering
\includegraphics[width=0.75\textwidth]{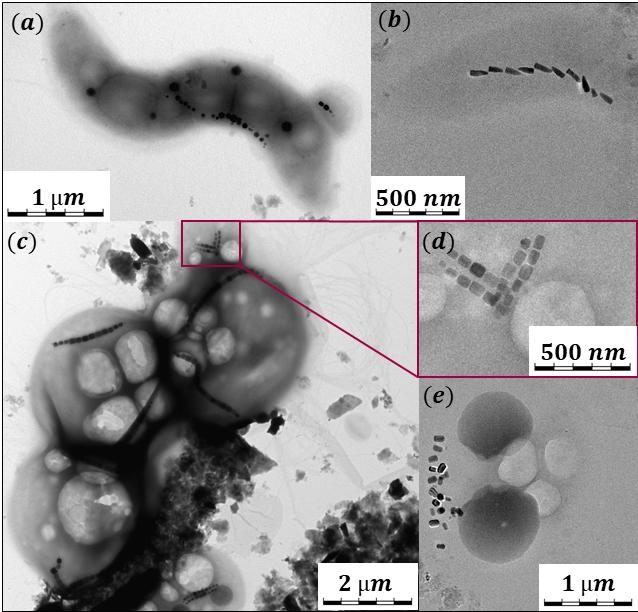} 
\caption{TEM images of diverse MTB found in the Ogre River: (a) a spirillum with a smaller MTB attached; (b) MTB with bullet-shape magnetosomes; (c) a cluster of cocci with 2 magnetosome chains in a single cell; (d) a double chain of prismatic magnetosomes; (e) a disordered magnetosome configuration.}
\label{fig:tem_mtb}
\end{figure}

The MTB found in the Ogre River show considerable diversity in cell morphology, as well as in magnetosome crystal shape and cellular localization and organization of crystals (Fig. \ref{fig:tem_mtb}). Different cocci and rod-shaped bacteria were observed in TEM. The clusters typically consist of 2 to 3 larger bacteria and several smaller ($d <1~ \mu m$) bacteria attached to the main cluster (Fig. \ref{fig:tem_mtb}c). TEM images show that some cocci clusters possess numerous flagella, although the clusters could be formed as artifacts of the TEM procedure.

Rod- and spiral-shaped bacteria of various sizes were found, and TEM images show that in some cases they possess a single flagellum. For some cocci, flagellar tufts could be seen in TEM, suggesting a bilopotrichous structure, also confirmed by the zigzag movement pattern characteristic of the helical motion \cite{lefevre_cultureindependent_2011} of some cells observed by optical microscopy (Fig. \ref{fig:cocci-track-and-image-examples}). 

The crystal shapes of magnetosomes in samples from the Ogre River also show considerable diversity: cubooctahedral magnetosomes (Fig. \ref{fig:tem_mtb}a), bullet-shaped (Fig.\ref{fig:tem_mtb}b), elongated prismatic (Fig.\ref{fig:tem_mtb}c,d) "ladyfinger"-shaped \cite{inproceedings} (Fig. \ref{fig:tem_mtb}e). Magnetosomes can be configured in a single chain per cell (Fig. \ref{fig:tem_mtb}a,b), two opposite chains in a cell (Fig. \ref{fig:tem_mtb}c), two double chains per cell (Fig. \ref{fig:tem_mtb}c,d) and a disordered cluster (Fig. \ref{fig:tem_mtb}e). Although some MTB sources can contain a single magnetic population, it is common to find mixed MTB species in an environmental sample \cite{xu_distribution_2018, flies_combined_2005, mtb_diversity_brazil} .

\begin{figure}[h]
\centering
\includegraphics[width=0.90\textwidth]{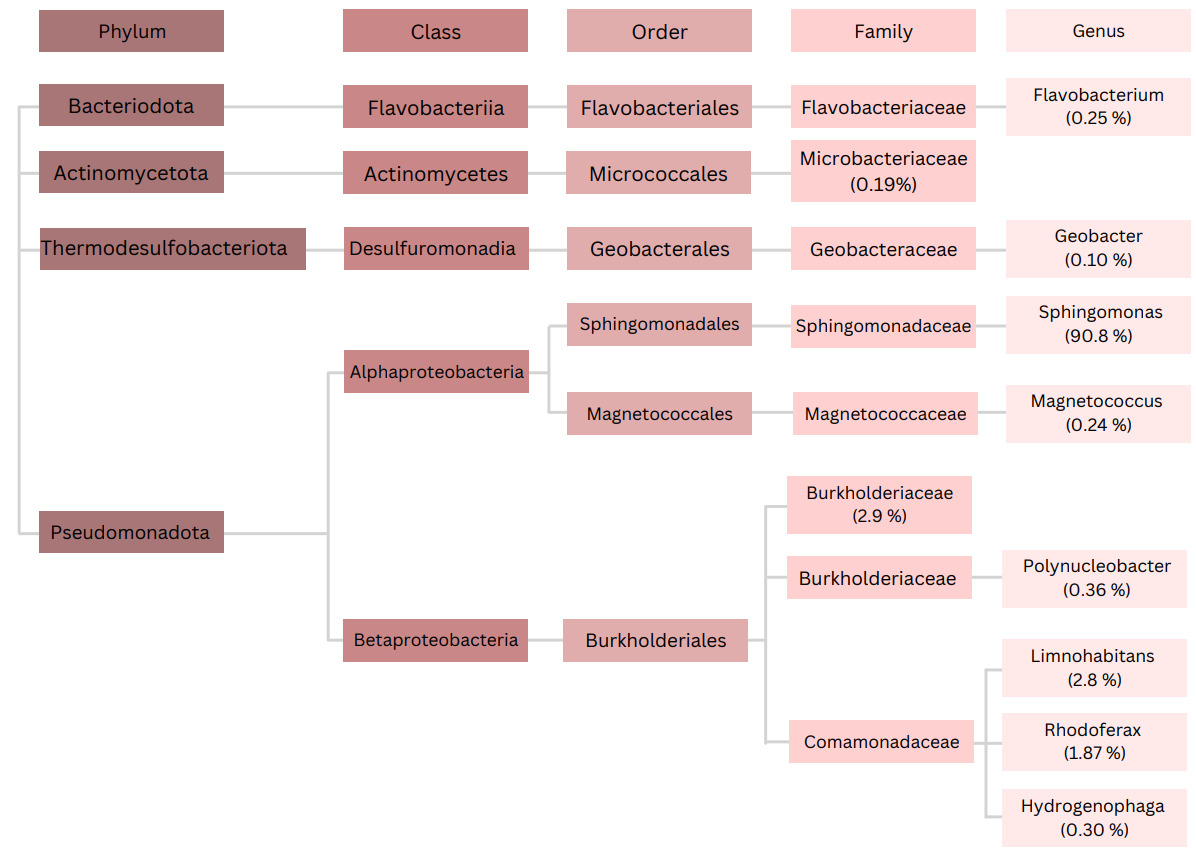} 
\caption{The relative abundance of bacteria found in Ogre River sample. MTB have been previously found in \textit{Sphingomona}, \textit{Magnetococcus} genera, as well as \textit{Burkholdericeae} family.}
\label{fig:Genetic analysis}
\end{figure}

16S rRNA metagenomic diversity analysis of the Ogre River sample was performed using the \textit{Illumina Miliseq} sequencing platform. The acquired data revealed a great diversity of microbiological organisms from eight families (Fig. \ref{fig:Genetic analysis}). The proteobacteria -alpha, -beta and -delta were present in the sample, but the dominant genus was \textit{Sphingomonas}, which amounted to $94.9\%$ of all sequenced bacteria. These were followed by $4.7\%$ of the sequencing reads coming from five different genera of the \textit{Burkholderiaceae} family, while $0.1\%$ of the read sequences were characteristic of the genera \textit{Magnetococcus} and \textit{Geobacter}. The genera \textit{Sphingomonas, Magnetococcaceae} as well as \textit{Burkholderiaceae} family are known to include MTB \cite{bazylinski_magnetococcus_2013}.  

Sphingomonas are typically associated with nutrient- and oxygen-rich media. Many species are associated with plant roots and potential also with soil effluxes to the river. However, there are results of the rich occurrence of Sphingomonas from deep sediments under Earth's surface in an oxygen-poor environment. Potentially, these anoxic strains convert ligninous substrates \cite{fredrickson_ecology_1999}. The appearance of Sphingomonas spp. in the magnetotactic enriched microbial community of the Ogre River may be related to the "contamination" of the river from rich soils of surrounding plants. In addition, it is possible that Sphingomonads exist permanently in the Ogre River, regardless of the soil \cite{Balkwill2006}. Interestingly, a similar pattern has been reported in the floodplain of the Araguaia River, Brazil, where most of the discovered MTB also belonged to the \textit{Pseudomonadota} phylum \cite{mtb_diversity_brazil}. 

\clearpage

\textit{Geobacter} bacteria are able to reduce Fe (III) in an anaerobic sedimentary environment, but are not known to form magnetosomes \cite{bazylinski_n_2000}. The metagenomic diversity confirms that the Ogre River is an environment where MTB can thrive, with an unknown number of species co-existing at the sampling site but belonging to at least three different families. 
As is typical in environmental samples, non-magnetic bacteria were also detected. Reads from \textit{Flavobacteriaceae} and \textit{Microbacteriaceae} families amounted to $0.1 \%$ of the entire repertoire each, but MTB have not yet been found in these genera, especially \textit{Microbacteriaceae}, as they are Gram positive and all MTB identified so far have been reported to be Gram negative \cite{SHIVELY2009404}.

\bigskip
\subsection*{Velocimetry-based characterization}
\label{sec:velocimetry}

\begin{figure}[H]
\centering
\includegraphics[width=1\textwidth]{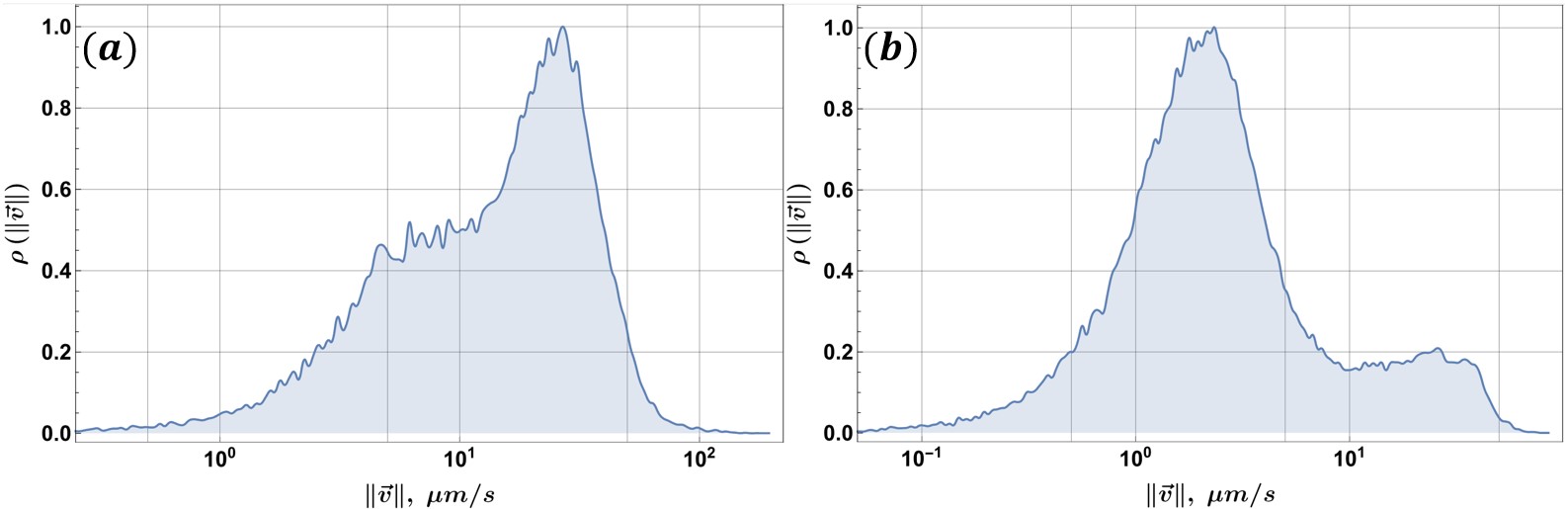} 
\caption{Relative frequency $\rho$ histogram for MTB velocity magnitude $\norm{\vec{v}}$ in the $\log_{10}$ scale: samples from Expedition B: (a) Day 1 and (b) Day 2. Freedman–Diaconis binning used in both cases.}
\label{fig:rel-velocities}
\end{figure}

\begin{figure}[h]
\centering
\includegraphics[width=0.75\textwidth]{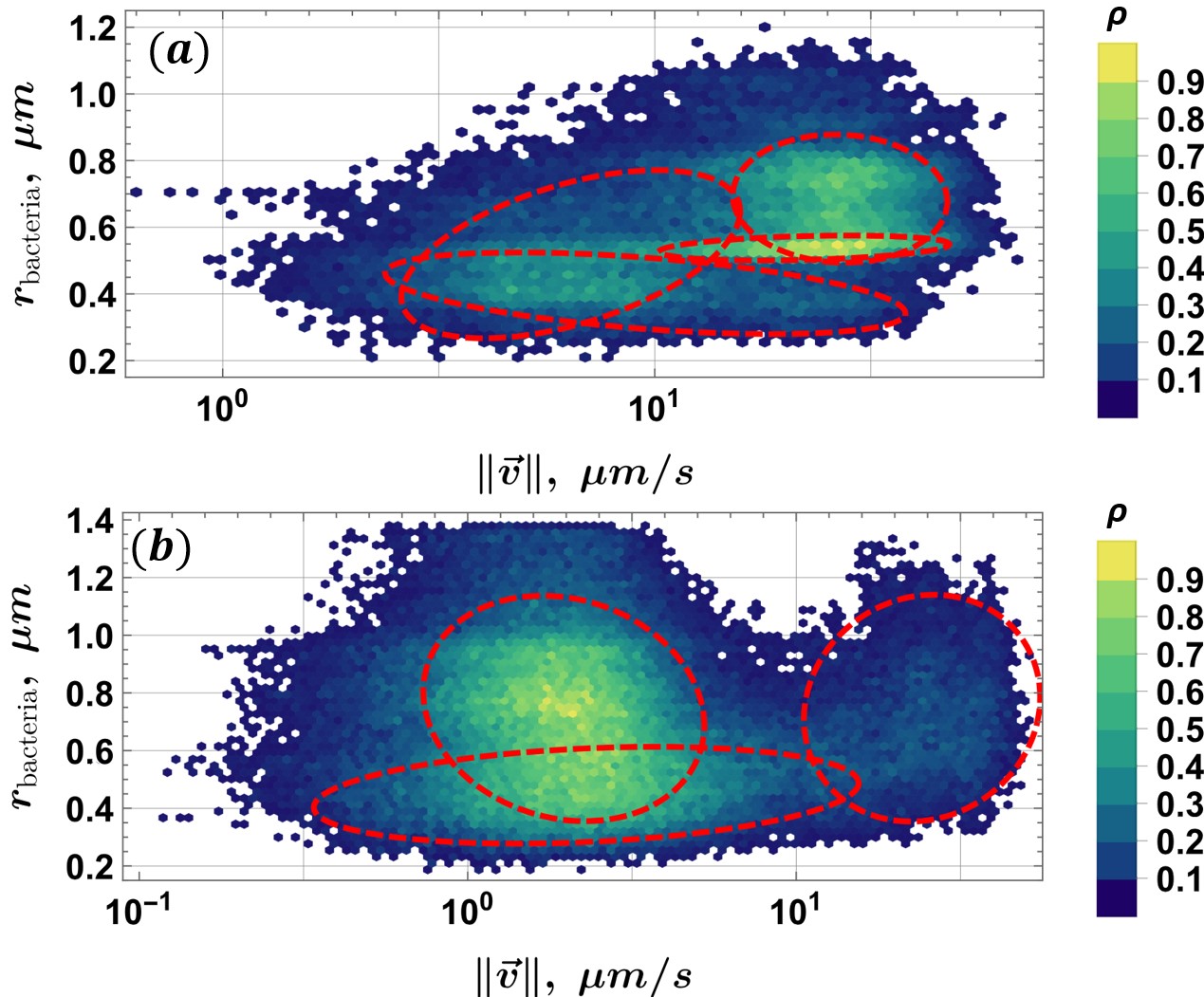} 
\caption{Relative frequency histograms for MTB effective radii $r_\text{bacteria}$ and velocity magnitude $\norm{\vec{v}}$ in the $\log_{10}$ scale, samples from Expedition B: (a) Day 1 and (b) Day 2. Freedman–Diaconis binning used in both cases. Dashed ellipses denote populations determined using the AIC-constrained multi-Gaussian fit, and the colour coding indicates the relative frequency.}
\label{fig:radii-velocity}
\end{figure}

\begin{figure}[h]
\centering
\includegraphics[width=0.75\textwidth]{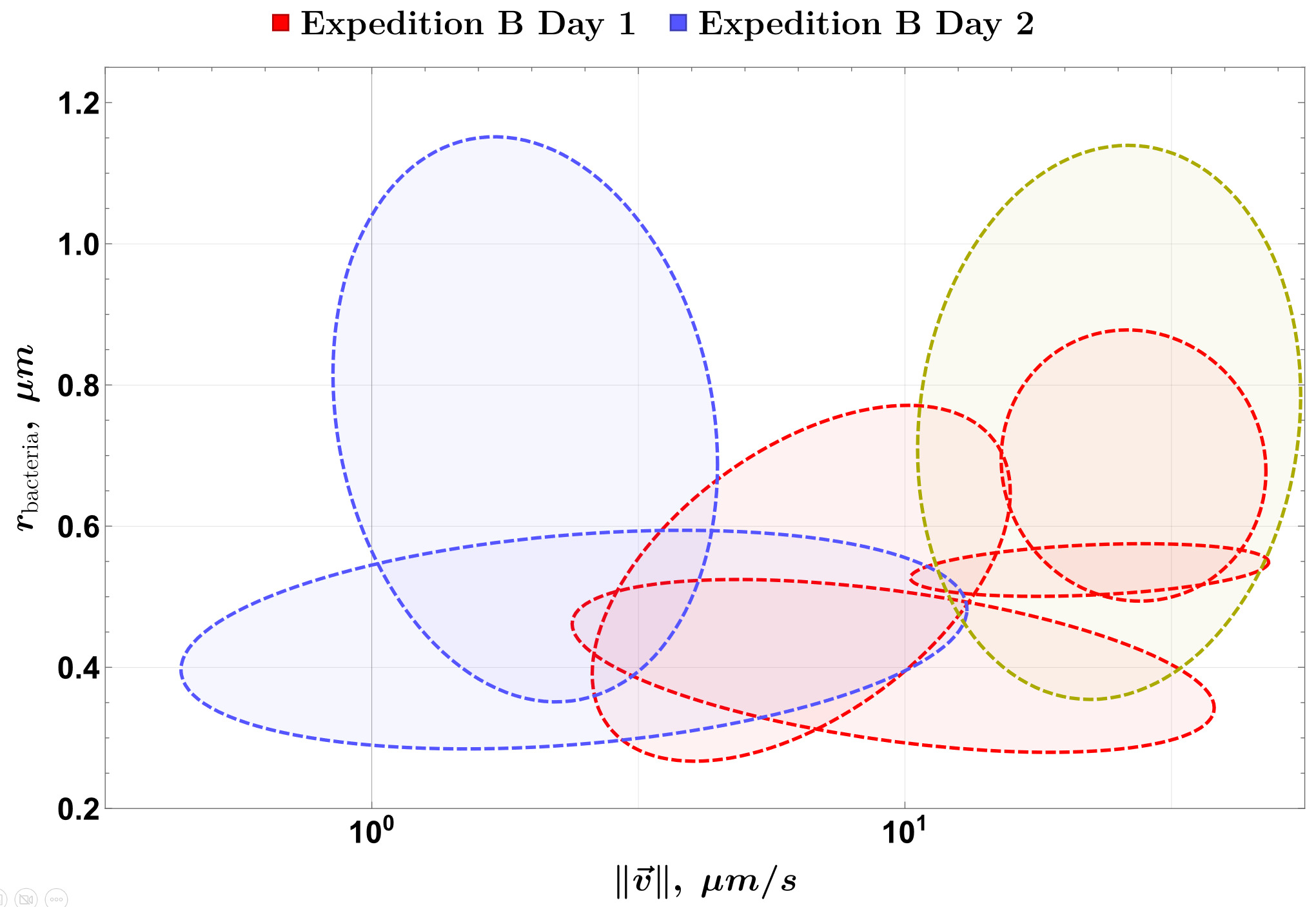} 
\caption{Population outlines for both measurement days (legend at the top). Dashed green line denotes one of the detected populations from Day 2 that is not significant.}
\label{fig:radii-velocity-populations-all}
\end{figure}

In addition to standard TEM and gene sequencing methods, we developed an open-source velocimetry data processing method (see the Code Availability section), based on conventional optical microscopy and a static magnetic field. In general, for data acquisition, this method only requires a microscope equipped with a camera and a small neodymium magnet as the MF source. The static magnetic field ensures that MTB are concentrated on the capillary boundary, and when it is reversed, MTB will follow the new MF direction. Long trajectories can then be recorded that span the whole field of view. 
The MTB were tracked, their velocities (Fig. \ref{fig:rel-velocities}) and effective radii were extracted using automated tools, and then multi-Gaussian fits, constrained by the Aikake information criterion (AIC) \cite{aikake-info-criterion}, were used to determine how many cell populations were present in the natural sample (Fig. \ref{fig:radii-velocity}; the population parameters are summarized in Tab. \ref{tab:coefficients_velocimetry_merged}; all populations are shown together in Fig. \ref{fig:radii-velocity-populations-all}). Note that these velocimetry results are not filtered MTB by compliance with applied MF direction, ensuring inclusion of all detected motion, but our code provides the option to filter out statistics for MTB not following the MF direction.

\clearpage

\clearpage

\begin{table}[H]
    \begin{center}
    \begin{tblr}{
      cells={valign=m, halign=c, mode=math},
      row{1}={bg=custompink},
      row{2,7}={mode=text, bg=customgray, font=\bfseries},
      colspec={|c|c|c|c|c|c|}
    }
        \hline
        \boldsymbol{\mu_1~ (\log_{10} \norm{\vec{v}},~\mu m/s)} & \boldsymbol{ \mu_2~ (r_\textbf{bacteria},~\mu m)} & \boldsymbol{\sigma_1} & \boldsymbol{\sigma_2} & \boldsymbol{\rho} & \norm{\vec{v}},~ \mu m/s \\
        \hline
        \SetCell[c=6]{} Day 1\\ 
        \hline
        0.98 \pm 0.01 & 0.402 \pm 0.002 & 0.40 \pm 0.01 & 0.081 \pm 0.002 & -0.49 \pm 0.02 & [3.8;24] \\
        \hline
        1.348 \pm 0.004 & 0.5379 \pm 0.0003 & 0.223 \pm 0.003 & 0.0247 \pm 0.0004 & 0.30 \pm 0.02 & [13;37] \\
        \hline
        1.429 \pm 0.002 & 0.686 \pm 0.001 & 0.165 \pm 0.001 & 0.127 \pm 0.001 & -0.046 \pm 0.012 & [18;39] \\
        \hline
        0.806 \pm 0.006 & 0.519 \pm 0.005 & 0.26 \pm 0.01 & 0.167 \pm 0.005 & 0.51 \pm 0.03 & [3.5;12] \\
        \hline
        \SetCell[c=6]{} Day 2\\ 
        \hline
        0.38 \pm 0.01 & 0.439 \pm 0.001 & 0.49 \pm 0.01 & 0.103 \pm 0.002 & 0.28 \pm 0.02 & [0.78;7.4] \\
        \hline
        0.288 \pm 0.001 & 0.751 \pm 0.003 & 0.239 \pm 0.001 & 0.265 \pm 0.02 & -0.16 \pm 0.01 & [1.1;3.4] \\
        \hline
    \end{tblr}
    \end{center}
    \caption{Population parameters for velocimetry, Days 1 and 2 (Figure \ref{fig:radii-velocity}), with uncertainties: $\mu_1$ -- mean velocity magnitude $\norm{\vec{v}}$ in $\log_{10}$ scale; $\mu_2$ -- mean bacteria radus; $\sigma_{1,2}$ -- standard deviations for $\mu_{1,2}$; $\rho$ -- correlation coefficient.}
    \label{tab:coefficients_velocimetry_merged}
\end{table}

\clearpage

The relative frequency velocity histogram from Expedition B (Fig. \ref{fig:rel-velocities}a) suggests that several populations may be present in the sample, with the most probable velocity $\sim 27~ \mu m/s$. Cell velocities up to $100~ \mu m/s$ are observed. The same velocimetry experiment using a fresh sample from the container was performed the next day (Fig. \ref{fig:rel-velocities}b) -- notice that the velocity distribution is now roughly bimodal, with the most probable velocity $\sim 2~ \mu m/s$ and another notable local maximum at $\sim 26~ \mu m/s$. The cell velocity range extends up to $\sim 50~ \mu m/s$, which means that faster cells were not present in the sample during the measurement or had perished in the 24-hour period since the previous experiment. Note also that the most probable velocity in (Fig. \ref{fig:rel-velocities}a) and the secondary maximum in (Fig. \ref{fig:rel-velocities}b) have consistent values, possibly corresponding to the same population that has diminished over time. Velocity distributions are a tool that is better suited for monoculture samples, as they can be difficult to interpret when multiple unknown distributions are present in the sample. For cases with an unknown number of populations, as in an environmental sample, evaluating a relative frequency histogram that relates cell velocity and radii (Fig. \ref{fig:radii-velocity}) could be more appropriate.

Figure \ref{fig:radii-velocity} shows the distribution of cells according to their effective radii $r$ and the corresponding velocity $v=\norm{\vec{v}}$. For multi-Gaussian fits AIC was used to penalize the parameter count of the Gaussian mixture model, and then only the populations that contribute significantly to the reconstruction of the integral volume of the probability density function (PDF) are selected as relevant. In Fig. \ref{fig:radii-velocity}a, four significant cell populations are found in the sample: two with a positive correlation $r(v)$ and two with a negative correlation. In contrast, in Fig. \ref{fig:radii-velocity}b only two populations remain, one with a positive correlation $r(v)$ and the other with a slightly negative correlation. The cell population with radii in the range between $r \in~\sim [0.2; 0.6]~ \mu m$ is present in the samples of both days.

The populations shown in Fig. \ref{fig:radii-velocity} were combined in a single plot for easier comparison in Fig. \ref{fig:radii-velocity-populations-all}. Two populations are steadily present on both days, but a third one, not statistically significant, denoted by a dashed green ellipse, appears on Day 2. Its area overlaps with the two populations that are present in the Day 1 sample, but not in Day 2. This could mean that it is a remnant of the two populations, where most of the cells have perished, but enough are still alive to be detected.

\clearpage

\subsection*{Magnetic moment-based characterization}
\label{sec:magnetic-moment}

\begin{figure}[H]
\centering
\includegraphics[width=0.75\textwidth]{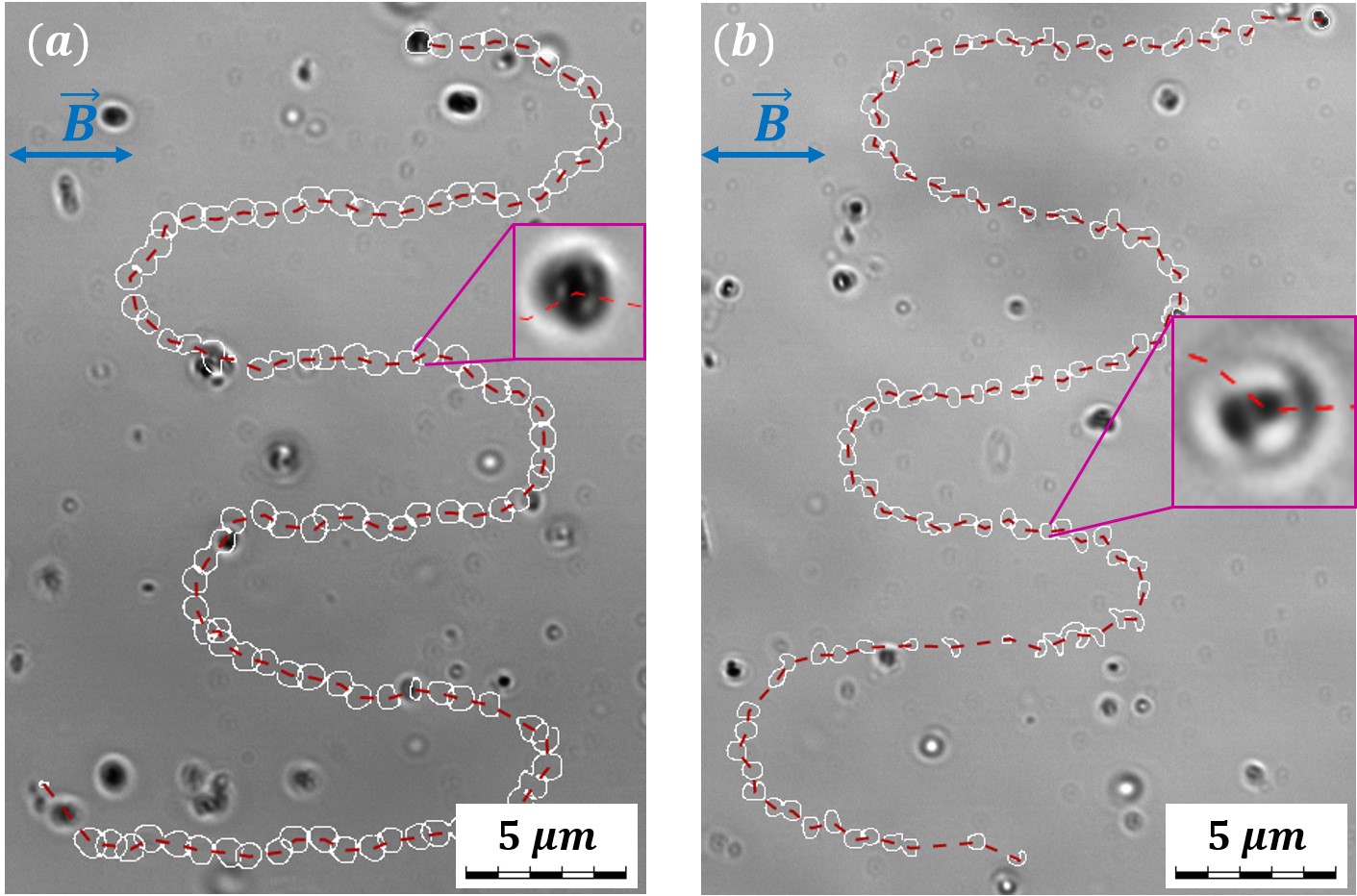} 
\caption{Examples of cell trajectories (red dashed lines) in an alternating MF: (a) a coccus and (b) a diplococcus. White contours outline the cell shapes.}
\label{fig:cocci-track-and-image-examples}
\end{figure}

Automatization of cell magnetic moment measurements is another powerful tool that allows us to categorize cells based on their magnetic properties, and give a deeper insight into cell behavior when used together with velocimetry. We used our automated U-turn method \cite{my_uturn_article}  to estimate the magnetic moment for 846 eligible individual tracks -- example trajectories are shown in Fig. \ref{fig:cocci-track-and-image-examples}. Cells from Expedition B Day 1 were studied.

\begin{figure}[H]
\centering
\includegraphics[width=0.75\textwidth]{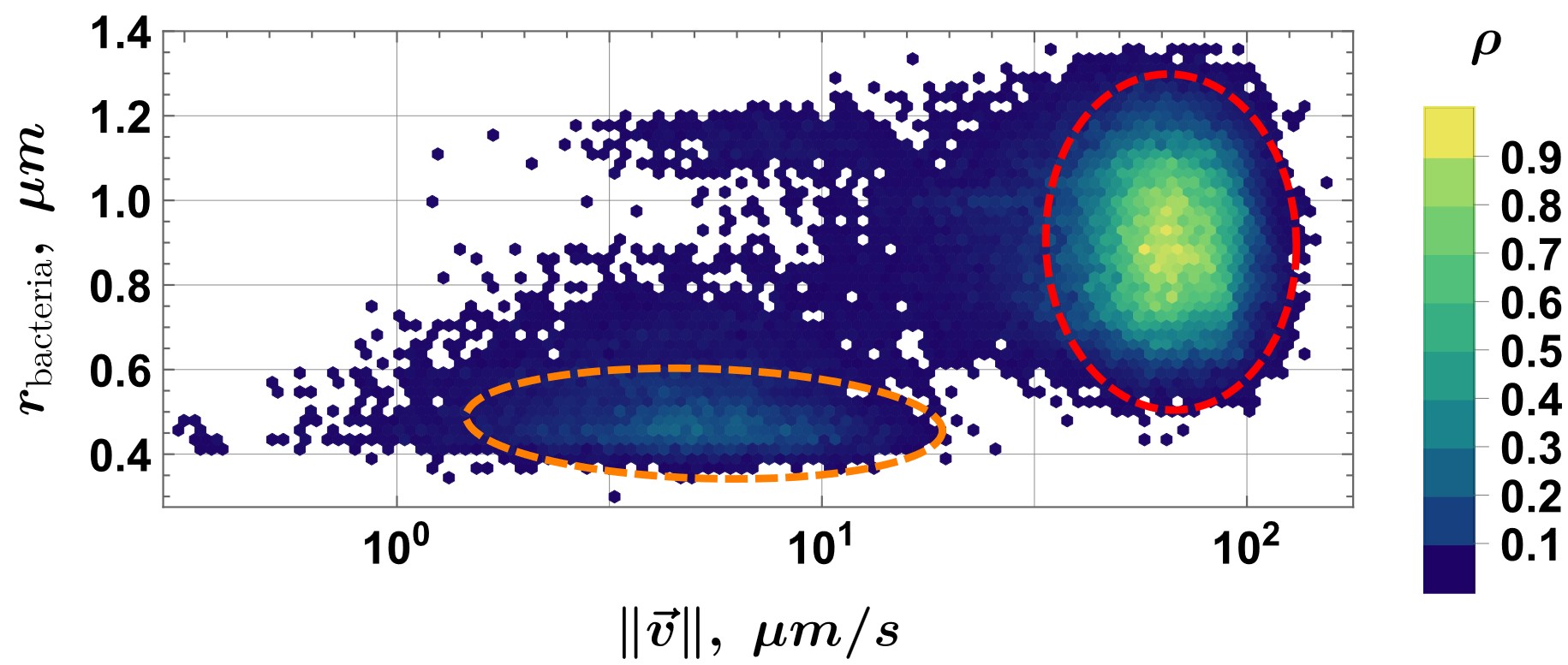} 
\caption{Data obtained from magnetic moment $m$ measurements via the modified U-turn method: (a) Velocitmetry: $v$ and effective radius $r$ distribution (Freedman–Diaconis binning), with significant populations marked with dashed ellipses -- orange colour for the nonmagnetic population, and red for magnetic population.}
\label{fig:moment_histogram_hexplot}
\end{figure}

\begin{figure}[H]
\centering
\includegraphics[width=0.70\textwidth]{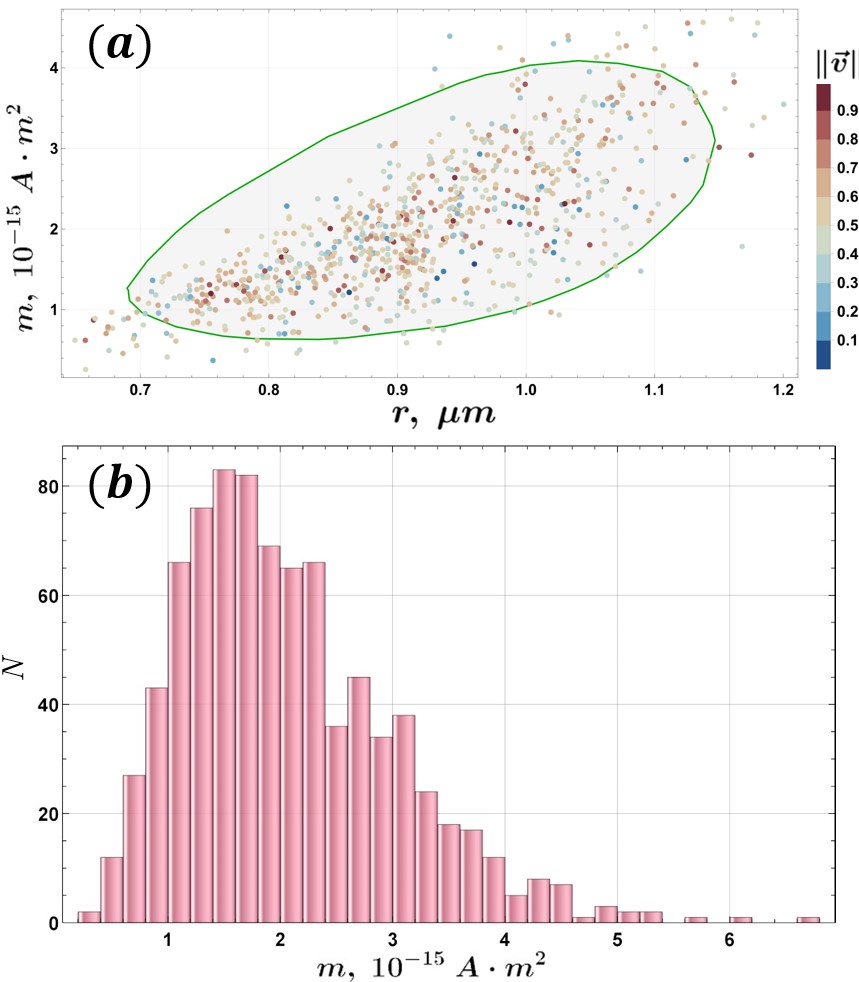} 
\caption{(a) Magnetic moment $m$ versus MTB effective radius. The measured $m$ values depending on the cell effective radius are represented by gray dots, with the $q=0.95$ quantile uncertainty region indicated as the light gray area with a green boundary. The uncertainty region is derived by computing 250 directional quantiles \cite{quantile-tomography, antonov-directional-quantile-envelopes}, connecting the quantiles with a smooth curve, and applying curve evolution polyline simplification \cite{polyline-simplification-curve-evolution} to the resulting quantile envelope. The data points are color coded by their normalized relative velocit; (b) magnetic moment histogram (count $N$ versus $m$, Freedman-Diaconis binning, $N =846$.)
}
\label{fig:r_m_tomography}
\end{figure}

The previously described velocimetry method was applied to the dataset, relating the MTB velocity and effective radius in a distribution shown in Fig. \ref{fig:moment_histogram_hexplot} two significant velocity/radius populations were found in the measured sample, and the magnetic moment $m$ distribution is shown in Fig. \ref{fig:r_m_tomography} b. The latter indicates that the most probable magnetic moment is $m=1.5 \cdot10^{-15} ~A\cdot m^2$, the mean magnetic moment is $\langle m \rangle=2.06\cdot10^{-15}~A\cdot m^2$ and its standard deviation is $\sigma (m) = 1.0\cdot10^{-15}~A\cdot m^2$. For comparison, the MSR-1 MTB $m \sim 10^{-16}~A\cdot m^2~$ \cite{zahn_measurement_2017, Pichel_2018, codutti_interplay_2022, my_uturn_article}, and for uncultured magnetotactic cocci the magnetic moment obtained by the U-turn method was reported as $m=8.2\cdot10^{-15}A~\cdot m^2~$    \cite{alvaros_u_turn_cocci}. $m \sim 10^{-13}~A\cdot m^2~$ has been reported for magnetotactic holobionts \cite{chevrier_collective_2023}. 

\clearpage

In Fig. \ref{fig:r_m_tomography} a, an overall positive correlation can be seen between the effective radius of the cells and their magnetic moment. A critical radius threshold is identified at $r_c = 0.57~ \mu m$ (an error-weighed linear model \cite{my_uturn_article}) below which the cell size is insufficient to host magnetosomes that produce measurable magnetic moment $m$. Similar positive correlations between bacteria size and magnetic moment have been reported in MSR-1 \cite{my_uturn_article} and in uncultured magnetic cocci \cite{alvaros_u_turn_cocci}.

\begin{figure}[h]
\centering
\includegraphics[width=0.85\textwidth]{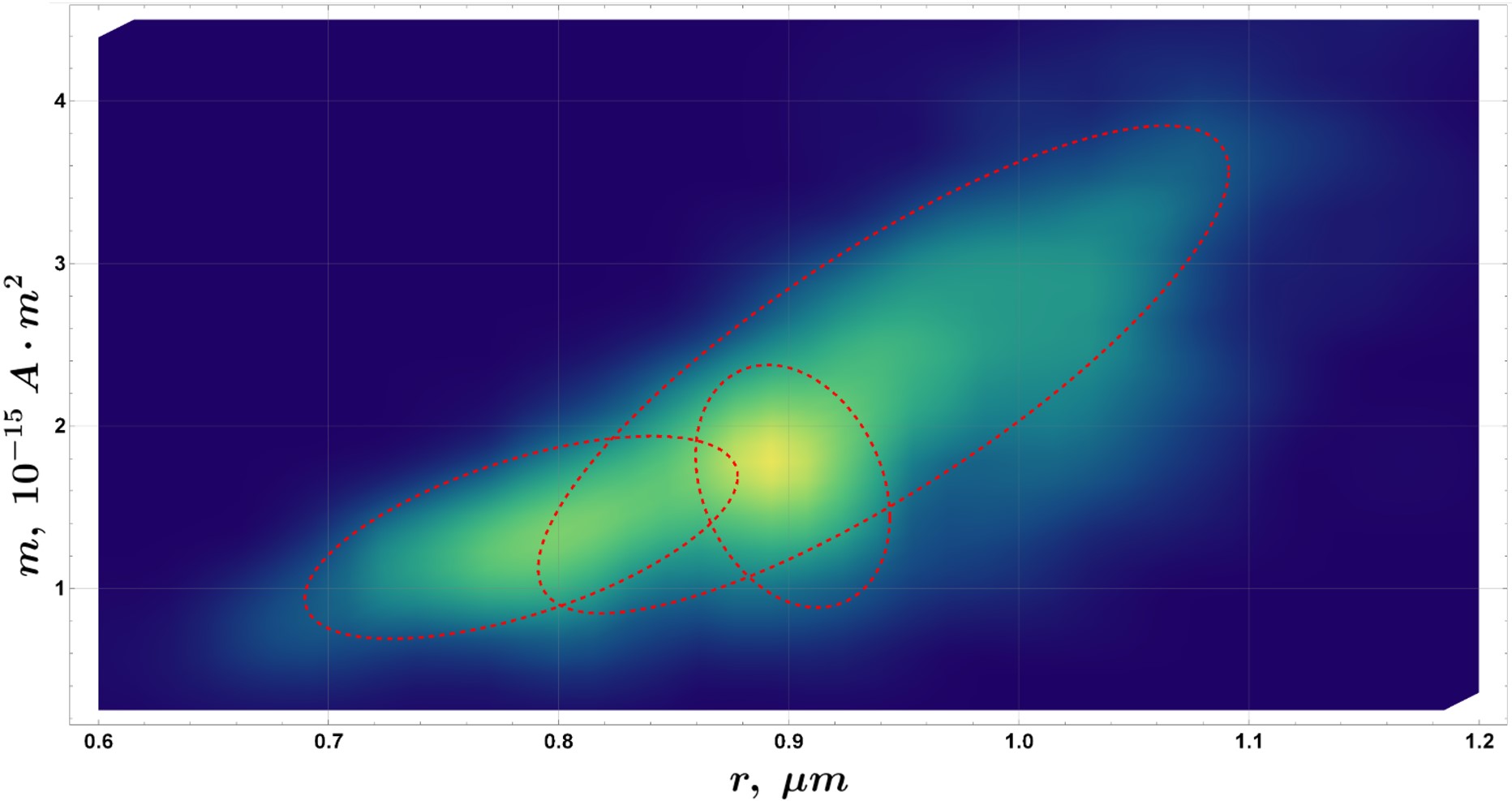} 
\caption{Magnetic moment $m$ versus MTB effective radius $r$: a smooth density histogram (Sheather-Jones bandwidth estimator, Epanechnikov kernel). Dashed ellipses represent AIC-constrained multi-Gaussian fits showing 3 significant populations present in the sample. For an extended analysis, please refer to \ref{appendix:A}.}
\label{fig:moment_histogram}
\end{figure}

A comparison between the velocimetry data (Fig. \ref{fig:moment_histogram_hexplot}a) and the smooth density histogram of magnetic moments (Fig. \ref{fig:moment_histogram}) reveals that the population of smaller bacteria detected in velocimetry is absent from the magnetic moment measurements. This discrepancy likely arises from the presence of non-magnetic bacteria in the sample, which can be tracked during motion analysis but do not exhibit U-turn behavior in response to alternating MF.

\clearpage

Data from the magnetic moment measurements can be further used to analyze the population composition of the sample. The analysis pipeline (please see the Code availability section) includes a built-in option to perform automatic Multi-Gaussian fitting of the magnetic moment versus cell effective radii smooth density histogram. As with velocimetry, the number of populations is restricted by AIC. Only statistically significant populations are presented in the resulting figure.

As shown in Fig. \ref{fig:moment_histogram}, the optimal fit according to the AIC corresponds to three populations, differentiated by their magnetic properties. Notably, one of the populations in this model exhibits a negative correlation between the cell radius and the magnetic moment, a trend that, to the best of our knowledge, has not been reported previously. It is unlikely that this correlation arises from cellular aggregates (e.g., doublets or triplets), as the corresponding effective radii ($0.86$–$0.93~ \mu m$) are too small to accommodate multiple cells. Thus, the presence of a distinct cluster cell population, as suggested in Fig. \ref{fig:tem_mtb}c, is not supported by the physical dimensions observed.

Furthermore, while holobionts have been reported to exhibit magnetic moments \cite{chevrier_collective_2023} on the order of $1.8 \pm 0.8 \cdot 10^{-13}~ A \cdot m^2$, such values are typically associated with protists that carry magnetotactic symbionts and have overall dimensions in the range of $10$-$20~ \mu m$. These dimensions are incompatible with the cell sizes observed in our study. Consequently, we interpret the three populations identified as representing three separate species, each represented at the single-cell level in the experimental analysis. 
It is important to note that a negative correlation is not set as a constraint in the Gaussian fitting process because it has not been strictly ruled out as impossible. The population parameters are shown in Tab. \ref{tab:coefficients_moment_hypothesis_2}.
\begin{table}[H]
    \begin{center}
    \begin{tblr}{
      cells={valign=m, halign=c, mode=math},
      row{1}={bg=custompink},
      colspec={|c|c|c|c|c|}
    }
        \hline
        \mu_1~ (m,~ 10^{-15}~ A \cdot m^2) & \mu_2~ (r_\text{bacteria},~ \mu m) & \sigma_1 & \sigma_2 & \rho \\
        \hline
        1.314 \pm 0.003 & 0.784 \pm 0.001 & 0.413 \pm 0.003 & 0.062 \pm 0.001 & 0.60 \pm 0.01 \\
        \hline
        1.63 \pm 0.01 & 0.9017 \pm 0.0003 & 0.49 \pm 0.01 & 0.0279 \pm 0.0003 & -0.26 \pm 0.02 \\
        \hline
        2.35 \pm 0.01 & 0.941 \pm 0.001 & 0.99 \pm 0.01 & 0.099 \pm 0.001 & 0.816 \pm 0.004 \\
        \hline
    \end{tblr}
    \end{center}
    \caption{Population parameters for the $m$ dataset (Figure \ref{fig:moment_histogram}).}
    \label{tab:coefficients_moment_hypothesis_2}
\end{table}

\clearpage

\section{Conclusions \& outlook}

We identified a previously undocumented source of MTB in the Ogre River, Latvia. Using a combination of TEM and 16S rRNA sequencing, and newly developed, automated image-based physical methods, we characterized the diversity and magnetic behavior of MTB in an environmental sample containing a mixture of MTB and non-magnetic bacteria. 

TEM images revealed a wide range of MTB morphologies, including cocci, rods, and spirilla, as well as diverse magnetosome crystal shapes: coboochtoedral, bullet-shaped, prismatic, and "ladyfinger"-shaped. Different magnetosome configurations were found: single, double, and multiple chains, as well as disordered structures. The presence of multiple MTB morphotypes was further supported by 16S rRNA analysis, which identified members of three families known to contain MTB: \textit{Sphingomonadaceae}, \textit{Burkholderiaceae}, and \textit{Magnetococcaceae}.

To analyze MTB behavior at the single-cell level, we developed and applied two open-source characterization methods: a low-equipment velocimetry pipeline based on standard optical microscopy and a static MF, and a modified U-turn method for estimating magnetic moments in an alternating MF. These methods enabled the identification and separation of distinct MTB populations based on cell size, swimming velocity, and magnetic moment. 

Velocimetry yielded velocity-radius histograms for two consecutive days, and combined with AIC-constrained multi-Gaussian fitting revealed that four distinct populations were present in the sample on Day 1. On Day 2, only two main populations remained, with overall reduced mobility. These results suggest that MTB from the Ogre River are highly sensitive to changes in the environment, which may explain their rapid decline under \textit{ex situ} conditions. Although velocimetry is a great tool for initial population estimation in a natural sample, it takes into account all moving bacteria, including non-magnetic ones, which may result in additional populations detected as a result. 

The U-turn method, while demanding a microscope equipped with an alternating MF generator, can be used to further distinguish between populations based on individual cell magnetic moments. We derived magnetic moment statistics from 846 trajectories, as well as put the data through the previously established velocimetry analysis pipeline. Although velocimetry identified two populations, analysis of the same dataset correlating magnetic moment and cell radius, and using AIC constrained multi-Gaussian fitting, found three MTB populations in the sample. One of the velocimetry populations was found to be nonmagnetic, bringing the total to four discernible groups.

Although AIC provides a statistically robust framework for model selection, it does not impose biophysical constraints and therefore does not prohibit populations with negative $m(r)$ correlations. In our analysis, we allowed for negative correlations, as the resulting three-population model provided the best fit according to AIC. Although such negative correlations would be biologically unusual, their presence in the data could reflect underlying species-level differences or uncharacterized magnetosome configurations. Another important consideration when performing population analysis of a wild sample is that different individual bacteria may have been recorded in each data set, due to the heterogeneous and uncharacterized composition of the wild microbial community. 

In general, this study demonstrates that the integration of automated velocimetry and magnetic moment analysis provides a powerful and accessible approach to differentiating between MTB populations in complex environmental samples.

\section*{Methods}

\subsection*{Sample collection}

Environmental samples were collected during two expeditions, the first (\textit{Expedition A}), conducted in August 2023, focused on sampling three freshwater sites to identify potential MTB habitats, while (\textit{Expedition B}) (October 2023) aimed to resample the single location in the Ogre River identified as a source of various MTB.
During Expedition A, 24 samples were collected in total, with at least seven from each freshwater body. Glass jars with metal lid were used as collection containers. The samples were obtained by manually filling the containers with a minimum of 1/4 of the sediment and the remaining volume with water. The sampling was carried out at depths up to 1 meter. The collected samples were stored at an ambient temperature of approximately $+20^\circ C$. Expedition B resulted in five samples collected under similar conditions. 

\bigskip

\subsection*{Bacteria extraction}

The extraction of magnetotactic bacteria was based on the method described by Lefevre and Bazylinski \cite{lefevre_cultureindependent_2011}. Before extraction, the contents of each container were thoroughly mixed by shaking in a circular motion. Subsequently, the container was placed between two $1.2$~ T magnets with the north and south poles oriented in opposite directions. The magnets were placed a few centimeters above the sediment line and away from the wall of the jar. The sample was left undisturbed for two hours, which did not yield results. Then it was mixed again and left to settle for 12 hours. 
If MTB are present in the sample, they tend to aggregate near magnets, forming a discernible pellet visible to the naked eye. The final step in the extraction process involved careful removal of the pellet from the inner side of the container with a pipette and its transfer to a $1.5~ ml$ microtube tube for further analysis. 

This technique yields approximately $30~ \mu l$ of concentrated bacteria suspension. Because this is a natural sample, a portion of the bacteria within the sample may not exhibit magnetic properties. Following the initial extraction of the pellet, the process can be repeated until pellet formation ceases, indicating the death or exhaustion of magnetic organisms in the sample. 

\bigskip

\subsection*{Magnetic property confirmation}

After extraction of the MTB, the magnetic properties of the sample were estimated using a Leica light microscope with $40\times$ magnification. For this purpose, microscopy slides were prepared by the following steps: a $14~mm$ diameter hole was cut into a double adhesive tape, which was then placed on a standard microscopy slide, thus creating a well-like structure. Several microliters of the bacterial suspension was pipetted into the center of the well, ensuring that the edges of the droplet did not touch the tape. The sample was then sealed with a glass cover slip. 

To assess the magnetic properties of the observed bacteria, we placed a weak permanent magnet on the microscope stage, with the south pole adjacent to the objective. Given that most bacteria are expected to be South-seeking, they would concentrate at the edge of the sample droplet closer to the magnet after a few minutes. The magnet was then flipped, positioning the north pole toward the objective, causing MTB to follow the South Pole and swim towards it. In contrast, if magnetic debris particles are present in the sample, they will rotate when the magnet is flipped, but they will not exhibit active swimming behavior, distinguishing them from motile microorganisms. 

As a result of the first expedition, we found that the location of the Ogre River is an abundant natural source of MTB. No MTB were found in the other two locations. 

\bigskip

\subsection*{Optical microscopy}

To analyze cell properties, a bacterial suspension was prepared using the previously described method. Measurements were performed with an optical microscope (\textit{Leica DMI3000B}) equipped with a $40 \times$ magnification objective. Images were recorded using a \textit{Basler AC1920155UM} camera, with a frame rate ranging from 20 to 100 frames per second. The cells were subjected to alternating $0.255~mT$ square wave MF. 

A magnetic coil system was designed and used to control the MF surrounding the sample. The MF was generated by three pairs of coils, arranged around a sample stage in a configuration that allows control of a homogeneous field in three dimensions. Each pair of coils was regulated via a data acquisition card and \textit{LabVIEW} code. The coils are powered by an AC power supply (\textit{Kepco BOP 2010M}).

\bigskip 
\subsection*{Image analysis}

An algorithm was developed which accounts for static and dynamic image artifacts, stuck and motionless MTB, field of view translation and (re-)focusing, uneven illumination, image noise, and bacteria shape metrics. The underlying methodology is detailed in \cite{my_uturn_article} and the code is open-source (see the Code Availability section).

\bigskip
\subsection*{Velocimetry}

To assess the velocity statistics for the sample, we applied a constant MF to the sample, and the motion of the bacteria was imaged using the light microscope. Using MF, cells were concentrated next to a wall of the capillary for several minutes, then MF was reversed, resulting in cell movement away from the wall. The recording continued until most of the cells had swum out of the field of vision. For both magnetic moment measurements and velocimetry, MTB trajectories were reconstructed using the MHT-X tracking algorithm \cite{mht-x-og, birjukovs-particle-EXIF, birjukovs-particle-track-curvature-stats}, which is open source (see the Code Availability section). Velocity and other track properties are readily derived from reconstructed tracks.

\bigskip
\subsection*{Magnetic moment calculation}

Cell magnetic moment measurement is performed via the standard U-turn method -- using a pulsed MF, alternating U-turn-like motion of bacteria is induced \cite{s_esquivel_motion_1986}. The U-turn width encodes the information about the MTB magnetic moment. To derive the magnetic moment statistics and scaling with MTB size from the acquired trajectories, we use our explicit and fully automatic method developed and previously used for MSR-1 MTB \cite{my_uturn_article}. Trajectories are first decomposed into U-turns, which are then fitted to a theoretical U-turn shape function derived in \cite{my_uturn_article}:

\begin{equation}
    y = - \frac{L}{\pi} \ln{ \left( \text{sec} \left( \frac{\pi x}{L} \right) \right) }
    \label{eq:uturn-theory-function}
\end{equation}
where $L$ is the asymptotic U-turn width, and $x$ and $y$ are the coordinates. When $L$ is determined, the cell magnetic moment $m$ can be determined as follows \cite{my_uturn_article}:

\begin{equation}
m = \frac{\pi \alpha v}{H L}
\label{eq:moment-from-uturn-params}
\end{equation}
where $\alpha$ is the MTB rotational drag coefficient with $\alpha = 8 \pi^2 \eta \cdot R^3$ , $\eta$ is the viscosity of the medium, $R$ is the effective radius of the microorganism, and $H$ is the strength of the applied pulsed MF. Here $\eta=0.90\cdot 10^{-3}~ Pa \cdot s$, the pulsed MF is set to $B= 2.55\cdot 10^{-4}~T$, and $R$ is determined for each bacterium as a trajectory average. For bacteria with irregular shapes, an effective radius (equivalent circle radius) was determined. The code for determining $m$ from MTB trajectories is open-source (please see the Code Availability section).

\bigskip
\subsection*{MTB population identification}

Population distributions are considered symmetric Gaussians $G(\mu_1,\mu_2,\sigma_1,\sigma_2,\rho,a)$ with means $\mu_{1,2}$, variances $\sigma_{1,2}$, covariance $\rho$, and scales $a$. An optimization sweep is performed where a different number of Gaussians $n \in N$ is used to fit the data by optimizing the parameters for all $G(\mu_1,\mu_2,\sigma_1,\sigma_2,\rho,a)$ using differential evolution. The optimal $n$ with the best fit is then given by the minimum Akiake information criterion (AIC) \cite{aikake-info-criterion}. The significant populations are then selected based on their contributions to the PDF integral volume reconstruction.

\bigskip
\subsection*{Transmission electron microscopy (TEM)}

 To characterize the morphology, as well as magnetosomes, transmission electron microscopy (TEM) was used. Samples were prepared by depositing bacterial suspensions, prepared using the method described in the "Bacteria extraction" subsection, onto carbon-coated copper TEM grids. The suspensions were then air-dried. The images were collected using a \textit{Tecnai G2 BioTWIN (FEI Company)} electron microscope equipped with a charged-coupled device (CCD) camera (\textit{Megaview III, Olympus Soft imaging Solutions GmbH}) using an accelerating voltage of $100~ kV$ and by a \textit{Talos F200X (Thermo Fisher)} TEM operated at $200~ kV$.

\bigskip
\subsection*{DNA extraction and sequencing of the 16S rRNA gene V3-V4 region}

    The DNA of the samples was isolated using \textit{FastDNA SPIN Kit for Soil (MP Biomedicals, USA)} according to the manufacturer’s guidelines. The quantity of the extracted DNA was assessed using \textit{Qubit dsDNA HS Assay Kit on the Qubit 2.0 Fluorometer (Thermo Fisher Scientific, USA)}.
    The two-stage PCR protocol was applied for \textit{MiSeq} library preparation. 341F and 805R primers were designed for the PCR amplification of the 16S rRNA gene V3-V4 region, which is specific to the domain \textit{bacteria}, and  contained Illumina overhang adapters \cite{fadrosh_improved_2014}. Microbial DNA (4 ng) was amplified separately using V3 and V4 primers using \textit{Phusion U Multiplex PCR Master Mix (Thermo Fisher Scientific) }under the following reaction conditions: denaturation at $98^\circ C$ for 30 seconds, 35 cycles of $98^\circ C$ for 10 seconds, $67^\circ C$ for 15 seconds, $72^\circ C$ for 15 seconds, and fragment elongation at $72^\circ C$ for 7 minutes. The yield of the acquired PCR products was assessed by 1. 2\% agarose gel electrophoresis and purified using a \textit{NucleoMag NGS Clean-Up and Size Select kit (Macherey-Nagel, Germany)}. The concentration of PCR products was measured using a \textit{Qubit dsDNA HS Assay Kit and a Qubit 2.0 Fluorometer}, and samples were normalized to $4~ ng/µl$. During the second PCR stage, \textit{Illumina MiSeq} i7 and i5 indexes were added to the $4~ ng$ of the V3 and V4 PCR product using custom-ordered \textit{Nextera XT Index Kit (Illumina Inc., USA)} primers (\textit{Metabion International AG, Germany}). For this reaction, \textit{Phusion U Multiplex PCR Master Mix} was used under the same thermal cycler reaction conditions as specified for the first PCR stage. The 16S rRNA PCR products were then pooled and purified for the sequencing reaction using \textit{NucleoMag} magnetic beads. The quality and acquired amount of the 16S rRNA V3-V4 amplicons were assessed using an \textit{Agilent High Sensitivity DNA Chip} kit and \textit{Agilent 2100 BioAnalyzer (Agilent Technologies, USA)} and a \textit{Qubit dsDNA HS Assay Kit and Qubit 2.0 Fluorometer}, respectively.
    Before sequencing, all samples were pooled at equal molarities and diluted to $6~ pM$. The samples were paired and sequenced using 500 cycles \textit{MiSeq Reagent Kit v2 on Illumina MiSeq (Illumina Inc., USA)}. After the sequencing run was completed, the individual sequence reads were filtered by \textit{MiSeq} software to remove low-quality sequences. All \textit{MiSeq} quality-approved, trimmed, and filtered data were exported as \textit{fastq} files.

\bigskip

\subsection*{16S rRNA gene sequencing data analysis}

    Sequence reads were quality filtered using \textit{Trimmomatic v.0.39} \cite{bolger_trimmomatic_2014} with the leading quality of Q20, the trailing quality of Q20 and sequences shorter than 36 nucleotides were discarded. All quality-approved sequences were imported into the \textit{QIIME 2 v.2021.11} \cite{bolyen_reproducible_2019} environment for further analysis. The \textit{DADA2} \cite{callahan_dada2_2016} plug-in was used to pair forward and reverse reads, as well as for additional sequence quality control and chimeric sequence removal using a pooled consensus method. The resulting feature table and sequences were used for the \textit{de novo} clustering employing \textit{VSEARCH} plug-in using the 97\% identity threshold \cite{rognes_vsearch_2016}. Thereafter, \textit{de novo} multiple sequence alignment was performed using the MAFFT method \cite{katoh_mafft_2013}, while phylogenetic trees were constructed using \textit{FastTree 2 }\cite{price_fasttree_2010}. \textit{De novo} clustered sequences were used for taxonomic assignment with the pre-fitted \textit{sklearn}-based \cite{pedregosa_scikit-learn_nodate} taxonomy classifier based on the \textit{SILVA v.132 97\%} identity reference database that was trained with the \textit{Naïve Bayes} classifier.

\section*{Appendices}

\subsection*{\textlabel{Appendix A}{appendix:A}: Population analysis in the $\mathbf{(r,v,m)}$ space}

\begin{figure}[h]
\centering
\includegraphics[width=0.7\textwidth]{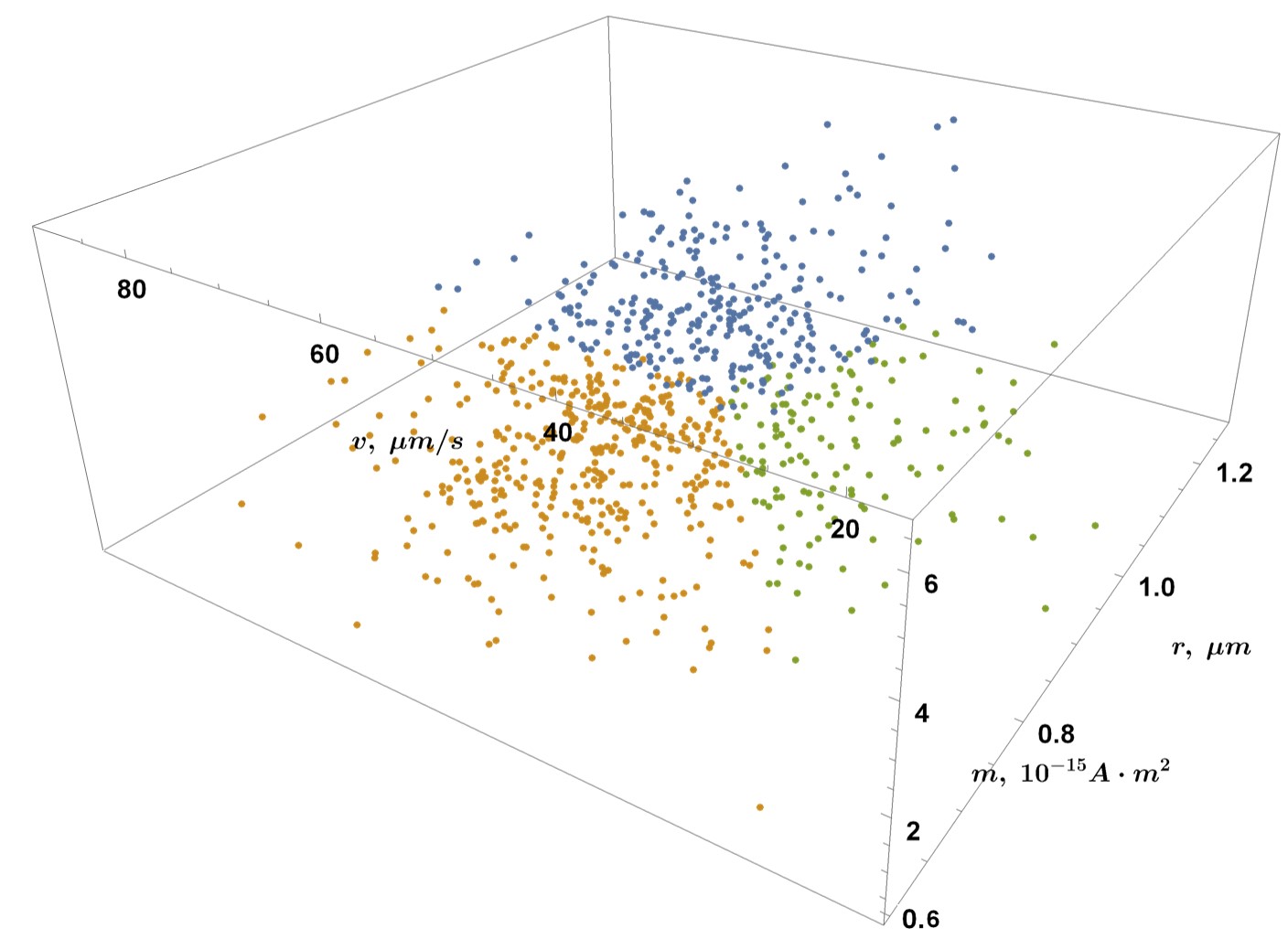} 
\caption{Population analysis in the $(r,v,m)$ MTB parameter space using variational Gaussian mixture clustering analysis. The three populations are colored in blue, yellow and green.}
\label{3d_correlation_moment_1}
\end{figure}

\begin{figure}[h]
\centering
\includegraphics[width=0.7\textwidth]{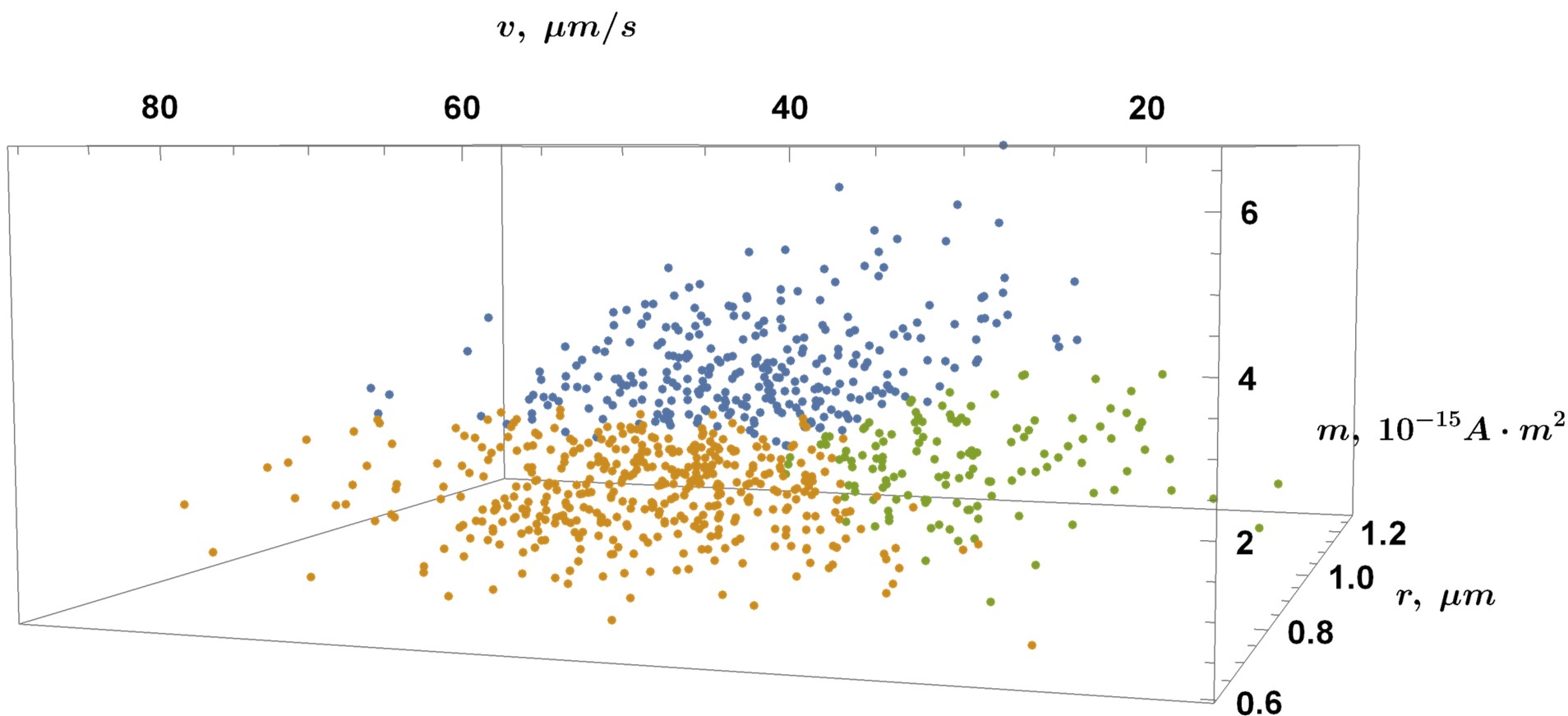} 
\caption{Population analysis in the $(r,v,m)$ parameter space, an additional point of view for clarity.}
\label{3d_correlation_moment_2}
\end{figure}

To further validate the three population hypothesis suggested by the 2D multi-Gaussian fitting (Fig. \ref{fig:moment_histogram}), a clustering analysis was performed for the magnetic moment dataset using a variational Gaussian mixture model \cite{gaussian-mixture-variational-ple} in the $(r,v,m)$ MTB parameter space. Although such clustering can provide insight into the population structure of the sample, we favor the 2D Gaussian linear combination constrained by AIC over $(r,m)$, as it accounts for possible population overlaps. In addition, in this case, there is not enough data for a 3D Gaussian linear combination fitting in the $(r,v,m)$ space, hence the strict clustering method for sparse data. As seen in Figs. \ref{3d_correlation_moment_1} and \ref{3d_correlation_moment_2}, three groups (colored blue, yellow, and green) can be distinguished based on the magnetic moment, radius, and velocity of the cell, similar to Fig. \ref{fig:moment_histogram}.

\clearpage

\begin{figure}[h]
\centering
\includegraphics[width=0.65\textwidth]{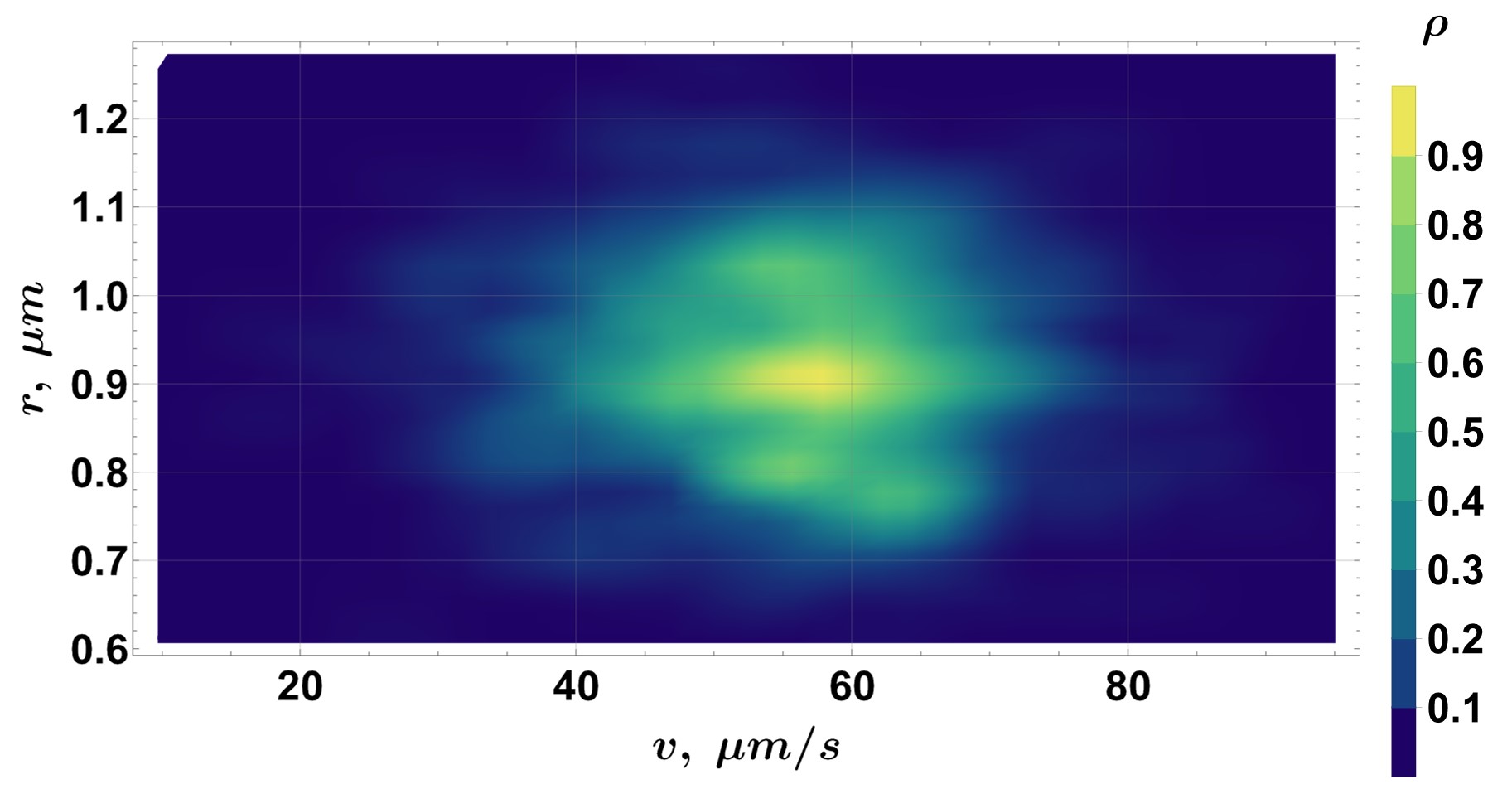} 
\caption{A smooth density histogram for the $(r,v)$ projection of the data in the $(r,v,m)$ space.}
\label{moment_smooth_dens_r_v}
\end{figure}

\begin{figure}[H]
\centering
\includegraphics[width=0.65\textwidth]{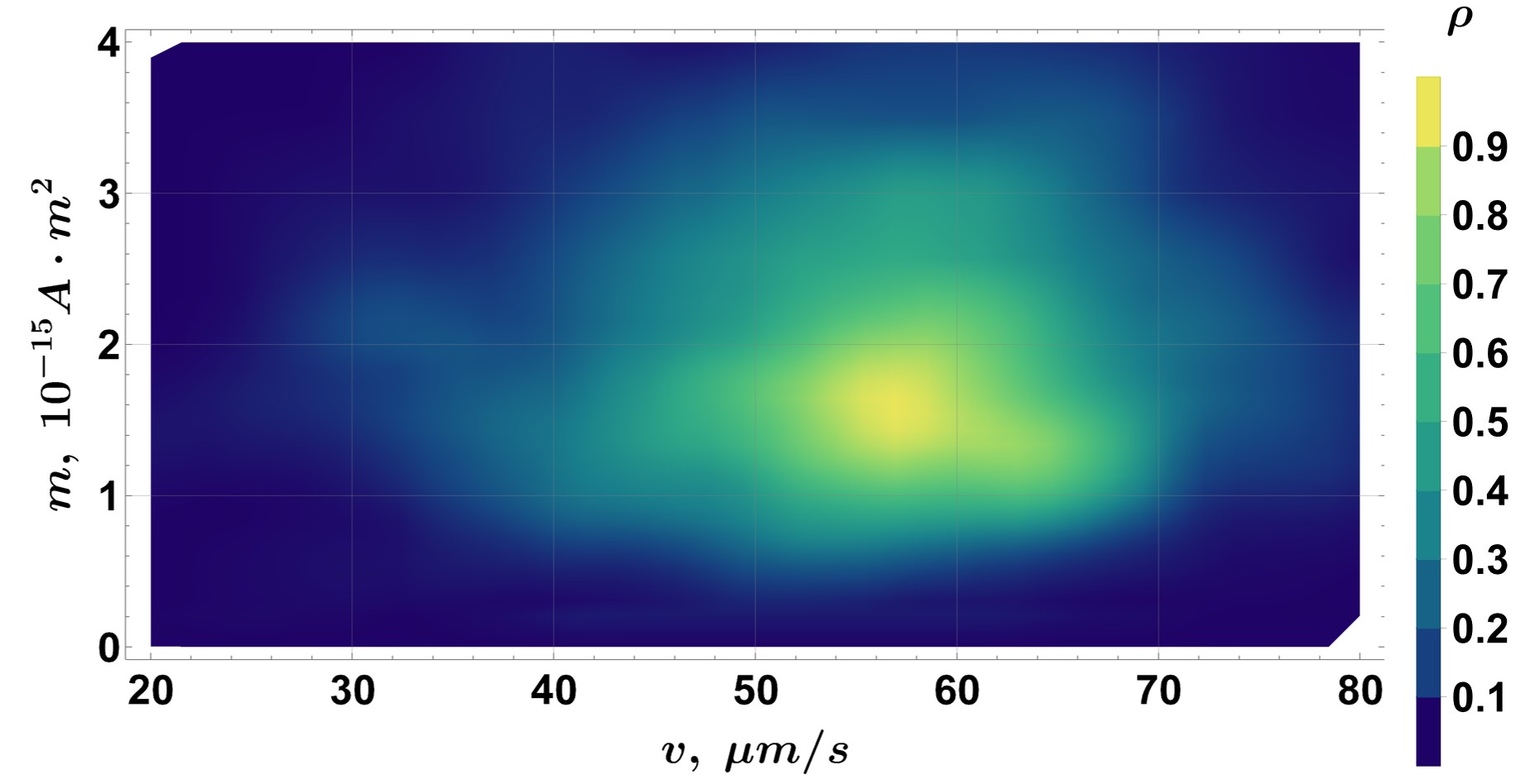} 
\caption{A smooth density histogram for the $(m,v)$ projection of the data in the $(r,v,m)$ space.}
\label{moment_smooth_dens_v_m}
\end{figure}

While a correlation between cell magnetic moment and its size has been shown here, it can also be informative to examine the smooth density histograms of the two $(r,v,m)$ data projections: $(r,v)$  (Fig. \ref{moment_smooth_dens_r_v}) and $(m,v)$ (Fig. \ref{moment_smooth_dens_v_m}). In Fig.\ref{moment_smooth_dens_r_v},  multiple groups can be distinguished according to their velocity and radius. Note that the global number density maximum in Fig. \ref{moment_smooth_dens_r_v} corresponds to the population with $\norm{\vec{v}} \in [18;39]~ \mu m/s$ and $r \in [0.56;0.81]~ \mu m$ in Fig. \ref{fig:radii-velocity}a (Table \ref{tab:coefficients_velocimetry_merged}), despite the bias introduced by the filtering criteria applied in the moment calculation algorithm. No correlation can be found between $m$ and $v$ in Fig. \ref{moment_smooth_dens_v_m}, as expected, as there is currently no evidence suggesting that the magnetic moment depends on the velocity of the cell.

\clearpage

\printbibliography

\section*{Acknowledgments}
Athors express their gratitude to Bhagyashri Shinde, Martins Klevs, Janis Cimurs and his children, and Malo Marmol for participating in the expedition to acquire bacteria samples. 
Mara Smite (M.S.), Mihails Birjukovs (M.B.), Andrejs Cebers (A.C.), and Guntars Kitenbergs (G.K.) acknowledge funding from the Latvian Council of Science, project A4Mswim, project Nr. lzp-2021/1-0470. 
M.S., Agnese Kokina (A.K.), Janis Liepins (J.L.), and G.K. acknowledge the LLC "MikroTik" donation project (no.2319), administered by the University of Latvia foundation, "Magnetotactic bacteria in nature and applications (MTBna)".
The authors are grateful to the French-Latvian bilateral program "Osmose", project Nr. LV-FR/2023/3. Damien Faivre (D.F.) acknowledges the BioMagnetLink project (grant agreement ID: 101187789). D.F. thanks M. FLoriani for access to the IRSN TEM.

\section*{Author contributions statement}
M.S. organized the expeditions for sample collection, extracted bacteria, performed optical microscopy, performed cell tracking experiments, initial magnetic moment estimates, interpreted results, and wrote the first draft; M.B. took part in sample collection, performed image processing and cell tracking, velocity measurements and magnetic moment calculation, data analysis and visualization, interpreted results, and wrote the first draft; Mila Kojadinovic-Sirinelli (M.K.-S.) participated in organization of the expedition, sample collection and bacteria extraction, performed field optical microscopy, prepared samples for TEM measurements, analyzed the results of TEM measurements and DNA sequencing; Sandrine Grosse (S.G.) participated in organization of the expedition, sample collection and bacteria extraction, performed field optical microscopy; A.K. took part in sample collection and facilitated field optical microscopy; J.L. provided research funding; Dita Gudra (D.G.) performed the 16S rRNA sequencing data analysis and contributed to the first draft; Megija Lunge (M.L.) performed DNA extraction and sequencing library preparation; Davids Fridmanis (D.Fr.) contributed to the manuscript editing and developed a strategy for sequencing experiments; Mihaly Posfai (M.P.) performed TEM measurements; A.C. took part in sample collection, provided research funding and supervised the study; D.F. contributed to the manuscript editing, performed TEM measurements and provided research funding; G.K. participated in organization of the expedition, contributed to the manuscript editing, provided funding, and supervised the study. All authors reviewed the manuscript prior to submission.

\section*{Code availability}
\label{sec:code_availability}

The image analysis, object tracking, and trajectory analysis code for $m$ calculations are available on \textit{GitHub}:

\begin{itemize}[noitemsep]
    \item MTB detection from images: \href{https://github.com/Mihails-Birjukovs/MTB_detection_tracking}{Mihails-Birjukovs/MTB\_detection\_tracking}
    \item MTB tracking: \href{https://github.com/Peteris-Zvejnieks/MHT-X/tree/particle_tracking_PIV_assisted}{Peteris-Zvejnieks/MHT-X}
    \item MTB magnetic moment retrieval from trajectories: \\ \href{https://github.com/Mihails-Birjukovs/MTB_magnetic_moment_from_U-turn_trajectories}{Mihails-Birjukovs/MTB\_magnetic\_moment\_from\_U-turn\_trajectories}
\end{itemize}

\end{document}